\documentstyle[aps,prb,psfig,floats,fancyheadings]{revtex}
\begin{document}
\twocolumn[\hsize\textwidth\columnwidth\hsize\csname@twocolumnfalse\endcsname
\title{Electronic structure of self-assembled quantum dots: comparison
between density functional theory and diffusion quantum Monte Carlo}
\author{J.~Shumway$^{a,*,\dag)}$, L.~R.~C. Fonseca$^{b,\ddag)}$,
J.~P.~Leburton$^{b,c)}$, Richard~M.~Martin$^{a,b)}$, D.~M.~Ceperley$^{a,b)}$}
\address{$^{a)}$Department of Physics \\
$^{b)}$Beckman Institute for Advanced Science and Technology \\
$^{c)}$Department of Electrical and Computer Engineering \\
University of Illinois at Urbana-Champaign, Urbana, IL 61801}
\date{\today }
\maketitle
\begin{abstract}
We have calculated the exchange, correlation, and total electronic energy 
of a realistic InAs self-assembled
quantum dot embedded in a GaAs matrix as a function of the
number of electrons in the dot.
The many-body interactions have been treated using 
the local spin density approximation (LSDA) to 
density functional theory (DFT) and diffusion
quantum Monte Carlo (DMC), so that we may quantify
the error introduced by LSDA.
The comparison shows that the LSDA errors are about 1-2 meV
per electron for the system considered.
These errors are small enough to justify the use of 
LSDA calculations to test models of self-assembled dots
against current experimental measurements.
\end{abstract}

]
\narrowtext

\section{Introduction}
The electronic structure of quantum dots resembles that of atoms.
Both systems display three dimensional
electronic confinement leading to level degeneracies, shell structure,
and spin correlation to only mention the most studied atomic
properties of quantum 
dots.~\cite{Frietal96,Taretal97,Drexler94,Marzin94,Grundmann95,Wojs96,Fonsetal98}
Direct as well as exchange Coulomb interactions are also present in quantum
dot systems, the importance of each depending on the dot
dimensions.~\cite{Fonsetal98} That is why some of the theoretical
methods used to calculate the electronic structure of quantum dots
are the same as the ones used to study atoms and molecules. One of
these methods, local spin density approximation (LSDA) 
density functional theory (DFT)
within the effective mass approximation (EMA), has been widely 
used to calculate the electron-electron interaction in
dots~\cite{Fonsetal98,Nagaraja97,Lee98,Macuetal97,Kosketal97}
because of its simple
implementation and low computer demand. LSDA is an
approximate theory, well known to
predict incorrectly the physical properties of some
molecules and solids, while performing well on many other
systems.\cite{Jones87} 
However, experience with LSDA on atoms and molecules
may not carry over to quantum dots since the confining potentials
and electron densities can be very different.
As a general rule, density functional theory tends to work better
for high effective density, and fails for low density systems, where 
correlation effects become important.  There have been several investigations
of LSDA and exact treatment of electron interactions in parabolic
quantum dots,\cite{Filippi94,Bolton96,Yang93,Eto97} with particular 
emphasis on low density systems and effects of external magnetic fields.
The electron interactions in parabolic dots have been treated exactly 
using diffusion Monte Carlo (DMC)\cite{Filippi94,Bolton96} as well as
exact diagonalization methods\cite{Yang93,Eto97}.
The parabolic potentials are popular models because they are computationally 
convenient and, with two adjustable parameters,
have had relative success in explaining the influence of many-body
effects on the electronic properties of dots.\cite{Wojs96}

In this paper we compare EMA results obtained with LSDA in self-assembled 
InAs/GaAs quantum dots against an exact treatment of the EMA many-body 
interactions obtained with diffusion quantum Monte Carlo.~\cite{CeKa79}
These small dots, which can hold about six electrons, have complicated
confining potentials that differ considerably from the parabolic dots
considered in previous DMC and exact diagonalization calculations.
The single particle states of realistically modeled 
pyramidal dots are significantly
modified from the states in a parabolic dot,~\cite{Fonsetal98} so 
many-body interactions should also differ from those in parabolic models.
The purpose of this comparison is to quantify errors due to LSDA for 
calculations on realistically modeled InAs/GaAs dots.
Although error of LSDA is expected to be small 
in these systems, it is large enough to be a limitation to 
the comparison of model results to experiments.  
For the system reported here, we consider LSDA errors of up to
10 meV to be acceptable given the current precision of the models and
experiments. Determination of whether LSDA meets this acceptability 
criterion requires careful comparison with many-body calculations
on realistically modeled potentials.

In these calculations we consider the ground state energy as a function of the 
number of electrons in the dot, which is the quantity
for which DFT should apply.  Electron addition and removal energies 
are experimentally measurable. These may be rigorously defined
as differences in the total energies of the ground states of the dot which
differ by one electron.  A common approach to calculating electron
addition and removal energies is to use the 
eigenvalues of the DFT equations,
which are well known not to be interpretable as electron addition and
removal energies.  We have addressed this 
issue in a previous paper~\cite{Fonsetal98}
where we have shown that addition and removal energies are to 
a very good approximation given by the ``Slater transition 
rule'',~\cite{Slater72}
which uses the eigenvalues at one-half occupation.  Thus the present work,
which compares total energies, is a rigorous test of the LSDA, and
combined with our previous work~\cite{Fonsetal98}
is a test of the accuracy of the addition and removal energies obtained using
LSDA eigenvalues and the Slater transition rules.

The ground state energy is the lowest eigenvalue
of the many-body Schr\"{o}dinger equation, which we take to be

\begin{equation}
\label{schro}
\left\{-\sum\limits_{i=1}^N
\frac{\hbar ^2}{2}\nabla_i \left[M^{-1}\nabla_i\right]+
V_{\rm cb}(R)+V_{\rm ee}(R)\right\}\Psi(R)=E\Psi(R),
\end{equation}
where $\Psi(R) = \Psi({\bf r}_1,{\bf r}_2,\cdots{\bf r}_N)$
is the $N$-electron wave function,
$M$ is the electron effective mass tensor and the potential
energy terms $V_{\rm cb}(R)$ and $V_{\rm ee}(R)$ are given by
\begin{eqnarray}
\label{potential}
V_{\rm cb}(R) &=& \sum\limits_{i=1}^N v_{\rm cb}({\bf r}_i) \\
V_{\rm ee}(R) &=& 
   \sum\limits_{i<j} \frac{e^2}{\epsilon | {\bf r}_i - {\bf r}_j |}
\end{eqnarray}
where $v_{\rm cb}({\bf r})$ is the potential 
for a single electron due to the offset and 
strain potential of the conduction band,~\cite{Fonsetal98} and
$V_{\rm ee}(R)$ is the Coulomb interaction of the conduction
electrons with charge $-e$ in a dielectric characterized
by the dielectric constant $\epsilon$.
The offset of the conduction band edge causes a step potential
at the edge of the dot, which is reduced by strain, 
resulting in a potential $v_{\rm cb}$ that is zero outside the dot and
a roughly constant depth of several tenths of an eV in the interior
of the dot, which typically has dimensions $\approx 10\,{\rm nm}$.
Strain causes the electron effective mass to become anisotropic leading
to a mass tensor given by 
${\rm diag}(M)=(m_{\rm xx} \; m_{\rm yy} \; m_{\rm zz})$
and zero off-diagonal terms.~\cite{Fonsetal98} In the usual case
of sample growth along the crystal direction (0 0 1), the electron
masses along the plane perpendicular to the growth direction are
equal, {\it i.e.} $m_{\rm xx}=m_{\rm yy}$.

\section{Density Function Theory and the Local Spin Density Approximation}
The DFT approach to obtaining the ground state energy 
is to replace the rather complex N-electron ground state 
wave function and the associated
Schr\"{o}dinger equation by the much simpler 
ground state electron density
$\rho({\bf r})$ and the corresponding functional forms $T[\rho]$
and $V[\rho]$ of the kinetic and potential energy operators,
respectively.~\cite{Hohenberg64,PaYa89,Dreizler90} 
However, those functional forms are 
unknown, and approximations are necessary. 
In Kohn-Sham theory\cite{Kohn76} the density is given by 
a sum of densities of single particle orbitals,
\begin{equation}
\label{density}
\rho({\bf r})=
\sum _\sigma \sum _{i=1}^N n_i|\psi _i({\bf r},\sigma)|^2.
\end{equation}
The total energy can then written as
\begin{equation}
\label{dft_energy}
E = E_{\rm kin} + E_{\rm cb}[\rho] + E_{\rm H}[\rho] + 
    E_{\rm x}[\rho] + E_{\rm c}[\rho],
\end{equation}
where
\begin{equation}
E_{\rm kin} =  -\frac{\hbar^2}{2}\sum\limits_{\sigma,i=1}^{N} \int 
    \psi_i^*({\bf r})\nabla (M^{-1} \nabla) 
    \psi_i({\bf r}, \sigma) d{\bf r}
\end{equation}
is the kinetic energy of the Kohn-Sham orbitals, 
\begin{equation}
E_{\rm cb}[\rho] = \int \rho({\bf r}) v_{\rm cb}({\bf r}) d{\bf r},
\end{equation}
is the potential energy from the conduction band offset and strain,
\begin{equation}
E_{\rm H}[\rho] = -e\int \rho({\bf r}) \phi({\bf r}) d{\bf r},
\end{equation}
is the Hartree energy, and $E_{\rm x}[\rho]$ 
and $E_{\rm c}[\rho]$ are 
functionals of the electron density.  The self-consistent electrostatic
potential $\phi({\bf r})$ is obtained as a solution to the 
Poisson equation
\begin{equation}
\label{poisson}
\nabla ^2 \phi({\bf r}) =\frac{e}{\epsilon}\rho({\bf r})
\end{equation}
with the boundary condition $\phi({\bf r}) \rightarrow 0$ as
$|{\mathbf r}| \rightarrow \infty$.
Minimization of the total energy defined by Eqs. (\ref{density} - 
\ref{poisson}) yields a one particle Schr\"{o}dinger equation 
which may be solved self-consistently with the expression for the density
and the Poisson equation to determine the exact
many-body ground state energy.  However, the nonlocal functionals
$E_{\rm x}[\rho]$ and $E_{\rm c}[\rho]$ are 
in general not known, so an approximation
must be made in order to proceed with the Kohn-Sham approach.

In the local spin density approximation (LSDA) the exchange and correlation
functionals are taken to be functions of the local charge
and spin density, which
are then uniquely defined by 
the requirement that the approximation be exact
for the homogeneous electron gas.  These functions have been fit 
to the exchange and correlation of the uniform electron gas.\cite{Perdew92}
The validity of LSDA has been a subject of much research in atomic,
crystalline, molecular, and parabolic quantum dot 
systems,~\cite{PaYa89,Dreizler90,Filippi94} but has not
been checked for self-assembled dot potentials, which are qualitatively
like finite square wells with rather high electron density.  
The density gradients and shell structures
of self-assembled dots cause the exchange and correlation energy to
differ from that of a uniform gas. The purpose of this paper is
to quantify this difference as a function of the number of electrons
in a realistically modeled dot.

\section{Quantum Monte Carlo Methods}
The simplest QMC method is variational
Monte Carlo (VMC), in which expectation values for a trial wave function
are evaluated using a Metropolis algorithm.  Using the variational
principle of the ground state energy, parameters in the wave function can be
optimized to minimize the total energy, thus
providing an approximation to the ground state energy and wave function.
In our calculation
we use VMC to optimize the guiding wave function 
for our diffusion Monte Carlo
algorithm, described below.

We also use Monte Carlo integration to evaluate the exact exchange energy.
In the Kohn-Sham formalism, the exchange energy is defined as the difference
\begin{equation}
E_{\rm x}[n] = \langle \Psi_{\rm KS} | V_{\rm ee} | 
\Psi_{\rm KS} \rangle - E_{\rm H}[\rho],
\end{equation}
where $| \Psi_{\rm KS} \rangle$ is a Slater 
determinant of the true Kohn-Sham orbitals; those that
give the exact density.
We assume that LSDA is accurate enough to provide approximations to the
Kohn-Sham orbitals, {\em i. e.} that the LSDA density is
close to the exact density.
We then evaluate the integral 
$\langle \Psi_{\rm KS} | V_{\rm ee} | \Psi_{\rm KS} \rangle$ using a 
Slater determinant of our calculated LSDA wave functions.

Diffusion Monte Carlo is a stochastic method that is able to 
project the N-electron ground state wave function $\Phi_0$
from a trial wave function $\Psi_{\rm T}$.
A position basis is used to described the state of the system.
A good trial wave function is important because the variance of the
Monte Carlo sampling decreases as the trial wave function approaches 
the true ground state wave function.  We chose the trial wave function
to be a product of Slater determinants for spin up and spin down electrons 
with a Jastrow factor that introduces correlation,
\begin{equation}
\Psi_{\rm T}(R) =\det [ \theta_{\rm k}({\bf r_{\rm i}}) ] 
     \times \exp [-\sum\limits_{\rm i<j} u(r_{\rm ij})].
\end{equation}
For convenience, we take the single particle wave functions 
$\theta_{\rm k}({\bf r})$ to be
the non-interacting eigenstates for the external potential,
and choose the Jastrow factor $u(r) = -ar/(1+br)$ with $a=m/\epsilon$
and $b=1/2$. 

Following the method described in Ref. [\onlinecite{CeKa79}], we
used importance sampling from a function 
$\Phi(R) = \Psi_{\rm T}(R) \Phi_0(R)$.
The anisotropic mass is taken into account using the mass tensor,
$M$.  The projection of a state $R$
to the ground state distribution $\Psi_{\rm T} 
\Phi_0$ is accomplished by
repeatedly applying the projection operator 
$e^{-H\tau}$, each time sampling
a new configuration.  Here $H$ is the Hamiltonian and $\tau$ is a 
parameter chosen to be small so that the 
projection operator is approximated by~\cite{CeKa79}
\begin{eqnarray}
\label{greenfunction}
\langle R|&&\Psi_{\rm T}(R) e^{-H\tau} \Psi_{\rm T}^{-1}(R') |R' \rangle 
\nonumber\\
&&=
    \frac{\exp\{-[E_{\rm L}(R)-E_0]\tau\}}
    {[(2\pi\tau)^3 \det M^{-1}]^\frac{n}{2}} \times\\
&&\phantom{=}
   \prod\limits_{i=1}^{n} \exp\left[-\frac{ 
        ({\mathbf r}_{\rm i} - {\mathbf r}_{\rm i} 
          - \tau {\mathbf v}_{\rm d,i}) 
        \cdot M \cdot
        ({\mathbf r}_{\rm i} - {\mathbf r}_{\rm i} 
          - \tau {\mathbf v}_{\rm d,i})} 
        {2 \tau }\right], \nonumber
\end{eqnarray}
where $E_0$ is a constant parameter to control the population,
$E_{\rm L}(R) = \Psi_{\rm T}(R)^{-1} H \Psi_{\rm T}(R)$ is the 
local energy, and 
${\mathbf v}_{\rm d} = M^{-1} \cdot \nabla \log \Psi_{\rm T}$.
The algorithm is thus: 
(1) Start with an ensemble of configurations (walkers)
distributed by $|\Psi_{\rm T}|^2$, (2) propagate each configuration
with a drift ${\mathbf v_{\rm d}}$ and 
Gaussian displacement with covariance
matrix $\sigma^2_{\rm ij} = M^{-1}_{\rm ij} \tau$, 
(3) reweight each configuration
by a factor $\exp\{-{(E_{\rm L}-E_0)} \tau\}$, 
(4) repeat, collecting statistics
once the steady state distribution is reached.  The exact 
ground state energy is obtained by sampling the local energy.
We use branching\cite{CeKa79} at each step to 
improve the efficiency of the process.

  A complication known as the fermion sign problem arises 
when applying diffusion Monte Carlo to fermions.
Electron exchange introduces negative signs into the
projection operator, which decreases the efficiency of the 
Monte Carlo sampling. In the present discussion 
the short time approximation, Eq. (\ref{greenfunction}), 
breaks down when walkers cross nodes of $\Psi_{\rm T}$. We
handle this problem with the fixed node approximation,~\cite{ReCe82}
in which walkers are given a weight of zero when they cross a node
of $\Psi_{\rm T}$. This approximation has a lower bound
property, so the true ground state energy can only be lower than
the fixed node energy, and the exact choice of 
$\Psi_{\rm T}=\Phi_0$ would give the exact Fermionic ground
state energy.

The correlation energy can be deduced from a knowledge of all
other energies,
\begin{equation}
E_c[\rho] = E_{\rm tot} - E_{\rm kin} - E_{\rm cb} - E_{\rm H} - E_{\rm x},
\end{equation}
or equivalently
\begin{equation}
\label{correlation}
E_c[\rho] =   \langle \Phi_0 | V_{\rm ee} | \Psi_0 \rangle
            - \langle \Psi_{\rm KS} | V_{\rm ee} | \Psi_{\rm KS} \rangle.
\end{equation}

\section{Numerical Comparison}
\begin{figure}[t]
\centerline{\psfig{file=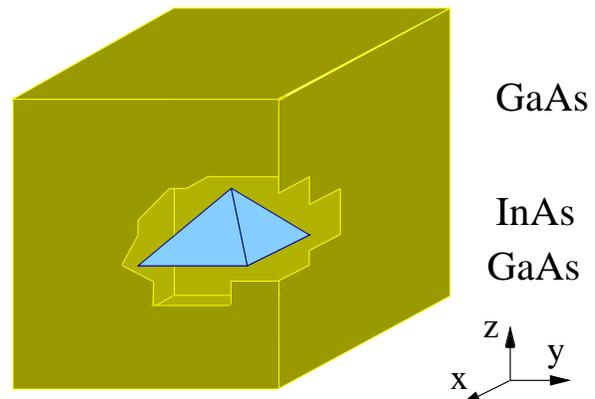,width=0.9\linewidth}}
\caption{Schematic representation of the self-assembled dot
structure used in the present work.
\label{device}}
\end{figure}
Using DFT and DMC we have calculated the total ground 
state energy of the system
of electrons, defined for DFT by Eq. \ref{dft_energy}, and defined 
within DMC as the fixed node ground state energy, as defined
in the previous section.
Figure~\ref{device} shows the system we have considered, 
a pyramidal shaped InAs quantum dot 
in a GaAs matrix.  However, our method of analysis is not restricted
to this particular shape, and we expect our results to be valid
for truncated pyramidal or lens shaped dots.
We have introduced several simplifications
in the model. Since our DMC scheme does not allow for a position
dependent dielectric constant, we have used only one value,
$\bar{\epsilon}=14\epsilon _0$, for both GaAs and for InAs.
In the absence of strain the GaAs - InAs conduction band offset 
between the two materials is 770 meV.
To partially take into account strain effects, 
we have assumed a constant shift of 400 meV in the InAs 
conduction band, and no strain effect in the GaAs
conduction band, resulting in a conduction band offset of 
$\rm 770 meV - 400 meV = 370 meV$.
We also take the electron effective masses to be constant in each
material.

In order to study a realistic potential, we have used a 
nonuniform grid
basis for the LSDA calculation and the solution of the Poisson equation.
For the Schr\"odinger equation we set the wave functions to zero
at the edge of the grid, which is reasonable due to the exponential
decay of bound states, but the Poisson equation requires more care.
We used a multipole expansion up to the quadrupole term to simulate
the boundary conditions at infinity.  This difficulty with
the boundary conditions of the Poisson equation is one source
of error in our comparison.
We used cubic interpolation 
to map the gridded wave functions and
conduction band potential to continuous functions for Monte Carlo 
evaluation of the exchange energy, for determination of the trial
nodes in the DMC calculation, and 
for guiding wave functions in the DMC calculation.
Although the cubic interpolation of the potential allows us to
compare LSDA and DMC for similar external potentials $v_{\rm cb}$,
errors in the kinetic energy operator are more difficult.
The use of a finite difference approximation to the Laplacian in the 
computation of the LSDA solution creates an operator that has no
simple counterpart in a continuous formulation of the problem.
The error between the physical Laplacian operator and the artificial
finite difference operator can be made arbitrarily small by the use
of finer grids, but computer memory and CPU 
time constraints caused this to be
a significant source of error in our comparison.
The size of the errors are several meV, which are comparable to the 
rather small LSDA errors, especially in the correlation energy.
Below we discuss a way we estimated this error so that we can compare
correlation energy. 

For the set of calculations we describe now, we have taken all masses
to have the isotropic value of $0.05 m_e$, for both the
InAs and GaAs regions.
The case of anisotropic masses will be discussed later. We have also
accounted for numerical discrepancies between the two calculations
due to grid interpolation error.
To estimate this error we have compared eigenvalues of single
electron states with both methods, and found that the DMC
single particle eigenvalues typically lie about 0.25 meV
above the eigenvalues computed by the grid method used in
the DFT calculation.  We attribute that difference to a
systematic error in the finite difference kinetic energy
operator and shift up our LSDA calculations to compensate for this.

\begin{figure}[t]
\centerline{\psfig{file=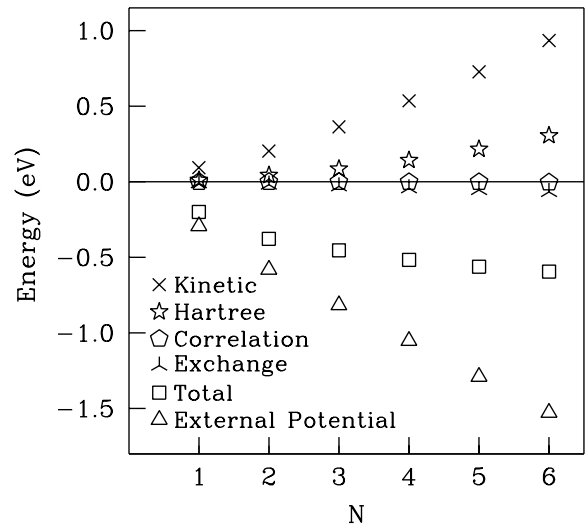,width=0.9\linewidth}}
\caption{Contributions to the energy of a $200 \times 100 {\rm \AA}$
pyramidal dot as a function of occupation $N$, calculated
by both LSDA and QMC.
Differences between LSDA and QMC are not apparent on the energy
scale of this figure.  This clearly shows the small effect of
interactions beyond the Hartree (mean field) level.
\label{energy_plot}}
\end{figure}
Figure~\ref{energy_plot} shows the values of 
the different components of the total
energy as a function of dot occupation.  Differences between LSDA
and QMC are not apparent on the scale of the figure.
This figure clearly shows that the external potential
energy and kinetic energy are much larger than the 
interaction energies.
In other words the effects of interactions enter as a perturbative 
correction to the non-interacting system.  The reason for this can
be seen from the scaled electron density.  The energy and length scales
for the electron interaction are scaled by the dielectric 
constant and mass, so
that the effective Bohr radius is $a_0^* = \epsilon/m^* a_0 
\approx 150 {\rm \AA}$ and the effective Hartree is
${\rm Ha^*} = \epsilon^{-2} m^* {\rm Ha} \approx 7 {\rm meV}$.
If we estimate the density by assuming the $N$ electrons uniformly 
occupy the interior of the dot, we find an effective 
conduction electron density of $r_s \approx 0.46 N^{-1/3}$.
To see the consequences of such a high effective density, consider
the uniform electron gas at $r_s = 0.6$.
The electron gas has a 
ground state energy expansion for small $r_s$,\cite{Mahan90}
\begin{equation}
\label{rs_expansion}
E = 2.2099 r_s^{-2} - 0.9163 r_s^{-1} -0.094 + 0.0622 \ln(r_s) +  \cdots,
\end{equation}
where the first term is the kinetic energy, the next term is the
exchange, and remaining terms are correlation energy.
For the case of six electrons in the dot $r_s \approx 0.25$ and the expansion
gives $E_{\rm kin} \approx 1440 {\rm meV}$, $E_{\rm x} \approx 150 {\rm meV}$, 
and $E_{\rm c} \approx 0.6 {\rm meV}$.
Although the comparison between
these very different electronic systems cannot be pushed too far, 
this does show that our exchange and correlation energies are reasonable
for this electron density.  The leading effect of the interactions
is the Hartree energy, with small corrections for exchange and very
small correlation corrections.

\begin{figure}[t]
\centerline{\psfig{file=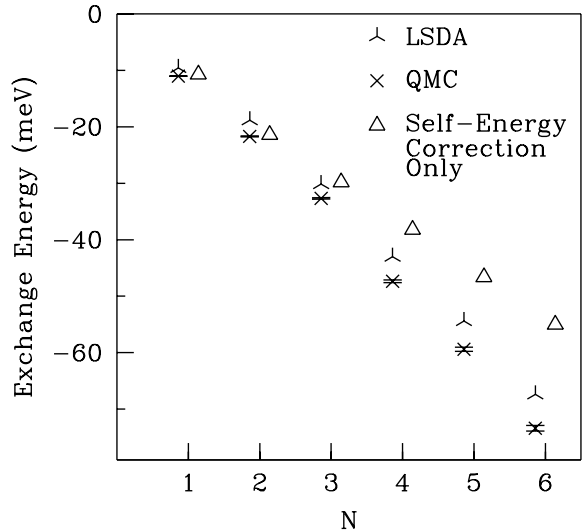,width=0.9\linewidth}}
\caption{Exchange energy of a $200 \times 100 {\rm \AA}$ pyramidal dot
as a function of occupation $N$, as determined by LSDA and QMC.
Also shown is the self-energy correction, which makes a large
contribution to the exchange energy.
\label{exchange_plot}}
\end{figure}
The most important contribution of LSDA in this system is the exchange 
term, and we plot the comparison of the LSDA exchange energy in
Figure~\ref{exchange_plot}.  
For finite systems, the definition for exchange includes
the correction for self-interaction in the 
Hartree energy,~\cite{PaYa89} which
we also show in the figure.  Merely correcting for self-interaction
will recover at least 75\% of the exchange energy, and
LSDA is able to do better, recovering about 90\%
for an error in the exchange energy of less than 5 meV.
As is well known in many other systems~\cite{PaYa89}, the error 
in the local approximation to the exchange is the largest error in LSDA.

\begin{figure}[t]
\centerline{\psfig{file=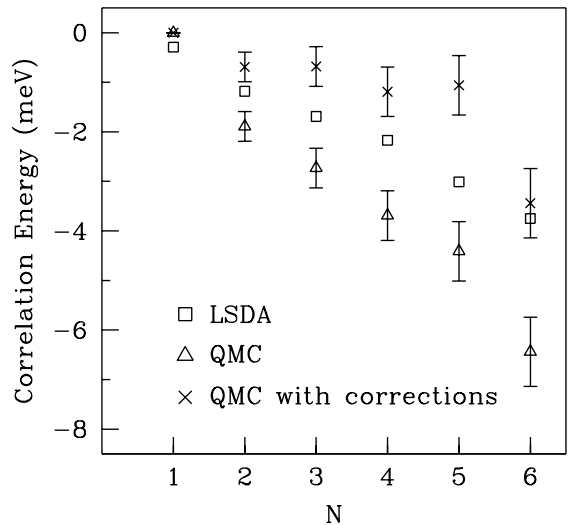,width=0.9\linewidth}}
\caption{Correlation energy of a $200 \times 100 {\rm \AA}$  pyramidal dot
as a function of occupation $N$, as determined by LSDA, QMC, and 
QMC with corrections for grid interpolation error, as explained in
the text.  LSDA gives a reasonable approximation to this small quantity.
\label{correlation_plot}}
\end{figure}
By comparison the correlation energy is much smaller
and its accuracy is difficult to assess due to grid errors.
Correlation is defined as the difference in total
energies, Eq. (\ref{correlation}),
and in this case the total energy difference is less than 1\% .  Thus our
estimates of the exact correlation energy are fairly uncertain.
In Figure~\ref{correlation_plot} we show our best attempt at checking
the LSDA correlation in the quantum dot.
Since the LSDA wave functions are not the true Kohn-Sham orbitals
(due mostly to grid errors) the correlation is
overestimated due to relaxation of our interpolated
LSDA states.  As stated earlier, we correct for this by
performing single particle calculations to estimate
the relaxation of our LSDA states,
and have removed this contribution from our calculated
correlation energy.  These results are shown as ``QMC'' and ``QMC
with corrections'' in the figure.  We find that LSDA gives a reasonable
estimate of the magnitude of the correlation energy in this system.

\begin{figure}[t]
\centerline{\psfig{file=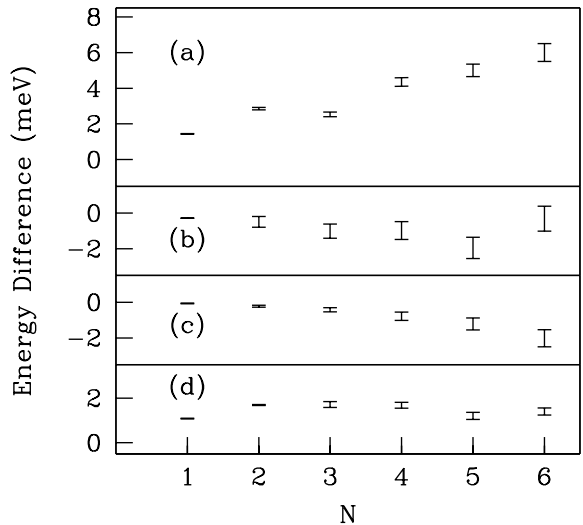,width=0.9\linewidth}}
\caption{Difference in energies between LSDA and QMC 
($E_{\rm LSDA} - E_{\rm DMC})$ for a 
$200 \times 100 {\rm \AA}$  pyramidal dot
as a function of occupation $N$:  
(a) error in the exchange energy, $E_{\rm x}$, due mostly to LSDA,
(b) error in correlation energy, $E_{\rm c}$, due 
to grid interpolation error and LSDA,
with some grid corrections as described in the text,
(c) error in Hartree energy $E_{\rm Ha}$,
due to grid interpolation error and the multipole expansion, and
(d) error in the total energy.  
The values in (a), (b) and (c) do not
add up to the value in (d) due to 
errors in $E_{\rm cb}$ and $E_{\rm kin}$ (not shown).
\label{en_diff_plot}}
\end{figure}
Our comparison of the energy errors from the LSDA 
are summarized in Figure \ref{en_diff_plot}.
The largest error is due to LSDA underestimating the magnitude
of the exchange energy, $E_{\rm x}$, 
which is a 6 meV error for six electrons in the dot.
There is a small error in the correlation energy, $E_{\rm c}$,
which slightly compensates for the error in the exchange.
The error in the Hartree energy $E_{\rm Ha}$ is due primarily to
truncation of the multipole expansion at quadrupole terms
for determination of the boundary conditions of the Poison equation.
There are also errors in $E_{\rm kin}$ and $E_{\rm cb}$ 
(not shown in the figure) due to the grid, in particular the
discretization of the kinetic energy operator discussed earlier.
We therefore find that LSDA leads to an error in the total
energy of up to 6 meV, which in our calculation is partially compensated
for by grid errors in other energy terms, leading to an overall 
error in the total energy of no more than 2 meV.

\begin{figure}[t]
\centerline{\psfig{file=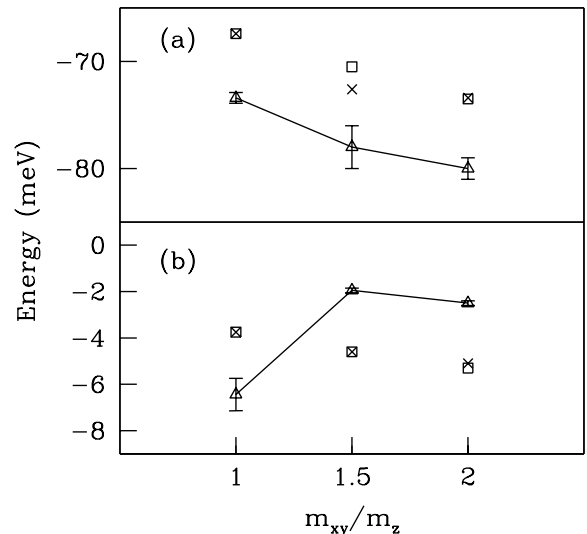,width=0.9\linewidth}}
\caption{
Values of the (a) exchange and (b) correlation energy as a function of
mass anisotropy for six electrons in 
a $200 \times {\rm \AA}$ pyramid dot.
Triangles are DMC results, $\times$'s are LSDA results with 
$m_{\rm xc}^{-1}= (2 m_\parallel^{-1} + m_\perp^{-1})/3$, and
squares are LSDA results with
$m_{\rm xc}=(m_{\parallel}^2*m_{\perp})^{1/3}$.
\label{m1_m2}}
\end{figure}
Because of strain, the electron effective mass in the plane parallel
to the base of the dot is usually different from the mass perpendicular
to it.~\cite{Fonsetal98}  
The LSDA does not explicitly account for mass anisotropy
in the exchange and correlation functionals, thus for
the calculation of the exchange-correlation potential $V_{\rm xc}$ we
have assumed a constant average mass given by
\begin{equation}
m_{\rm xc}=(m_{\parallel}^2 m_{\perp})^{1/3},
\end{equation}
where $m_{\parallel}$ is the in-plane mass, $m_{\parallel}=m_{\rm xx}=
m_{\rm yy}$, and $m_{\perp}$ is the perpendicular mass, $m_{\perp}=
m_{\rm zz}$. The choice of the average mass $m_{\rm xc}$ 
is somewhat arbitrary.
We have performed a test of this approximation by calculating the
exchange and correlation energy for an anisotropic system using
(i) LSDA with $m_{\rm xc}$ as given above, (ii) LSDA with $m_{\rm xc}$ 
optically averaged, $m_{\rm xc}^{-1}= (2/m_\parallel + 1/m_\perp)/3$,
and (iii) DMC, which can explicitly treat anisotropic mass.
Figure~\ref{m1_m2} shows the calculated exchange and correlation energies
for several electron mass ratios, which we have again fixed to constant
across the InAs and GaAs.  We find that either choice is acceptable,
as both give errors comparable to the isotropic mass case.

\section{Conclusions}
In conclusion, we have shown that DFT offers an accurate
approximation for the ground state energy 
including many-body interactions in small self-assembled
quantum dots, while providing a simple and fast means of modeling
systems containing several electrons.  This is in large part due to the
small size of the self-assembled quantum dots in relation to the effective
Bohr radius.
In this regime the largest error is in the local approximation to the
exchange, and we have verified that the errors due to this approximation
are small.  This is quantified in Figures~\ref{energy_plot} 
and \ref{exchange_plot} where we see
that the exchange energy for six electrons in the dot is 72 meV
with an LSDA error of 6 meV,
compared to a total energy of 1490 meV. 
We expect this applicability of LSDA to hold in general 
for small quantum dots of various shapes, but we emphasize that LSDA 
is expected to become progressively worse if the dot
becomes much larger than the effective Bohr radius. 
Also, the use of the EMA Schr\"odinger equation,
Eq. (\ref{schro}), although reasonable for the chosen problem,
is predicted to fail for very small dots.~\cite{FrZu97}
Although our model assumes pure InAs dots surrounded by
GaAs, in some systems indium and gallium may in fact intermix.~\cite{Sk00}
Again, we expect the relative magnetude of exchange and correlation
energies and the acceptability of LSDA to be unchanged by
intermixing or alloying of the dots.
 
Based on this comparison we conclude that the 
addition and removal energies found by 
EMA-LSDA calculations in realistic
dots~\cite{Fonsetal98} are accurate to 1-2 meV per
electron as far as the many-body
interactions within the EMA are concerned.  Thus, comparison
of such calculated charging energies
with experiment can be considered a direct test of the models 
and the uncertainties in the analysis of 
experiments.~\cite{Frietal96,Fonsetal98}  Furthermore, we
have carefully tested the grid errors and concluded that
energies are accurate to $\approx \pm 10$ meV including all errors.

We would like to point out that application of LSDA to
a system of several coupled self-assembled dots may 
have difficulties.  Systems that have weak intersite
transitions along with intrasite interactions are
a well-known many-body problem where LSDA is known to fail.
The problem is closely related to the $\rm H_2$ molecule,
where LSDA describes well the electronic energy at equilibrium
distance, but gives an incorrect broken symmetry solution 
if the atoms are pulled apart beyond a critical distance.
The problem of weakly coupled dots would be of particular
importance for future studies.

\section*{Acknowledgments}
This work was supported by CRI from the University of Illinois
and NSF Grants No. ECS 95-09751 and No. DMR 94-22496 and computer
resources at NCSA.

\bibliographystyle{prsty}

\end{document}